Editors' Suggestion Featured in Physics

# Experimental Observation of Proton Bunch Modulation in a Plasma at Varying Plasma Densities

E. Adli,[1] A. Ahuja,[2] O. Apsimon,[3,4] R. Apsimon,[5,4] A.-M. Bachmann,[2,6,7] D. Barrientos,[2] M. M. Barros,[2] J. Batkiewicz,[2] F. Batsch,[2,6,7] J. Bauche,[2] V. K. Berglyd Olsen,[1] M. Bernardini,[2] B. Biskup,[2] A. Boccardi,[2] T. Bogey,[2] T. Bohl,[2] C. Bracco,[2] F. Braunmüller,[6] S. Burger,[2] G. Burt,[5,4] S. Bustamante,[2] B. Buttenschön,[8] A. Caldwell,[6] M. Cascella,[9] J. Chappell,[9] E. Chevallay,[2] M. Chung,[10] D. Cooke,[9] H. Damerau,[2] L. Deacon,[9] L. H. Deubner,[11] A. Dexter,[5,4] S. Doebert,[2] J. Farmer,[12] V. N. Fedosseev,[2] G. Fior,[6] R. Fiorito,[13,4] R. A. Fonseca,[14] F. Friebel,[2] L. Garolfi,[2] S. Gessner,[2] I. Gorgisyan,[2] A. A. Gorn,[15,16] E. Granados,[2] O. Grulke,[8,17] E. Gschwendtner,[2] A. Guerrero,[2] J. Hansen,[2] A. Helm,[18] J. R. Henderson,[5,4] C. Hessler,[2] W. Hofle,[2] M. Hüther,[6] M. Ibison,[13,4] L. Jensen,[2] S. Jolly,[9] F. Keeble,[9] S.-Y. Kim,[10] F. Kraus,[19] T. Lefevre,[2] G. LeGodec,[2] Y. Li,[3,4] S. Liu,[20] N. Lopes,[18] K. V. Lotov,[15,16] L. Maricalva Brun,[2] M. Martyanov,[6] S. Mazzoni,[2] D. Medina Godoy,[2] V. A. Minakov,[15,16] J. Mitchell,[5,4] J. C. Molendijk,[2] R. Mompo,[2] J. T. Moody,[6] M. Moreira,[18,2] P. Muggli,[6,2] C. Mutin,[2] E. Öz,[6] E. Ozturk,[2] C. Pasquino,[2] A. Pardons,[2] F. Peña Asmus,[6,7] K. Pepitone,[2] A. Perera,[13,4] A. Petrenko,[2,15] S. Pitman,[5,4] G. Plyushchev,[2] A. Pukhov,[12] S. Rey,[2] K. Rieger,[6,*] H. Ruhl,[21] J. S. Schmidt,[2] I. A. Shalimova,[16,22] E. Shaposhnikova,[2] P. Sherwood,[9] L. O. Silva,[18] L. Soby,[2] A. P. Sosedkin,[15,16] R. Speroni,[2] R. I. Spitsyn,[15,16] P. V. Tuev,[15,16] M. Turner,[2] F. Velotti,[2] L. Verra,[2,23] V. A. Verzilov,[20] J. Vieira,[18] H. Vincke,[2] C. P. Welsch,[13,4] B. Williamson,[3,4] M. Wing,[9] B. Woolley,[2] and G. Xia[3,4]

(AWAKE Collaboration)

[1]*University of Oslo, 0316 Oslo, Norway*
[2]*CERN, 1211 Geneva, Switzerland*
[3]*University of Manchester, M13 9PL Manchester, United Kingdom*
[4]*Cockcroft Institute, WA4 4AD Daresbury, United Kingdom*
[5]*Lancaster University, LA1 4YB Lancaster, United Kingdom*
[6]*Max Planck Institute for Physics, 80805 Munich, Germany*
[7]*Technical University Munich, 80333 Munich, Germany*
[8]*Max Planck Institute for Plasma Physics, 17491 Greifswald, Germany*
[9]*UCL, WC1E 6BT London, United Kingdom*
[10]*UNIST, 44919 Ulsan, Republic of Korea*
[11]*Philipps—Universität Marburg, 35032 Marburg, Germany*
[12]*Heinrich-Heine-University of Düsseldorf, 40225 Düsseldorf, Germany*
[13]*University of Liverpool, L69 7ZE Liverpool, United Kingdom*
[14]*ISCTE—Instituto Universitéario de Lisboa, 1649-026 Lisbon, Portugal*
[15]*Budker Institute of Nuclear Physics SB RAS, 630090 Novosibirsk, Russia*
[16]*Novosibirsk State University, 630090 Novosibirsk, Russia*
[17]*Technical University of Denmark, 2800 Lyngby, Denmark*
[18]*GoLP/Instituto de Plasmas e Fusão Nuclear, Instituto Superior Técnico, Universidade de Lisboa, 1049-001 Lisbon, Portugal*
[19]*Philipps-Universität Marburg, 35032 Marburg, Germany*
[20]*TRIUMF, V6T 2A3 Vancouver, Canada*
[21]*Ludwig-Maximilians-Universität, 80539 Munich, Germany*
[22]*Institute of Computational Mathematics and Mathematical Geophysics SB RAS, 630090 Novosibirsk, Russia*
[23]*University of Milan, 20122 Milan, Italy*

We give direct experimental evidence for the observation of the full transverse self-modulation of a long, relativistic proton bunch propagating through a dense plasma. The bunch exits the plasma with a periodic density modulation resulting from radial wakefield effects. We show that the modulation is seeded by a relativistic ionization front created using an intense laser pulse copropagating with the proton bunch. The modulation extends over the length of the proton bunch following the seed point. By varying the plasma density over one order of magnitude, we show that the modulation frequency scales with the expected dependence on the plasma density, i.e., it is equal to the plasma frequency, as expected from theory.

*Introduction.*—Recent plasma wakefield experiments have demonstrated very high accelerating fields acting on electrons. The accelerating field in a plasma with electron density $n_{pe}$ can reach a significant fraction of the wave breaking field $E_{wb} = (m_e c/e)\omega_{pe}$, where $\omega_{pe} = [(n_{pe}e^2)/(\epsilon_0 m_e)]^{1/2}$ is the angular plasma electron frequency [1]. This field is $> 1$ GV/m for plasma densities $>10^{14}$ cm$^{-3}$ ($E_{wb}[\mathrm{V/m}] \cong 100\sqrt{n_{pe}[\mathrm{cm}^{-3}]}$), which makes plasma a promising candidate as a medium for high-gradient acceleration. Plasma wakefields can be driven by intense laser pulses [2] and relativistic charged particle bunches [3] that can be compressed to subpicosecond durations. To effectively drive wakefields, the drive bunch rms length $\sigma_z$ must be on the order of the cold plasma skin depth such that $\sigma_z/(c/\omega_{pe}) \cong \sqrt{2}$ [4]. Electrons have been used as drive bunch [5,6]. Numerical simulations using a single, short ($\sim$100 $\mu$m) drive bunch and a single plasma stage [7] suggest that high-energy proton bunches can be used to drive wakefields and accelerate electrons to the TeV energy scale. Proton bunches are typically long: $\sigma_z = 6$–$12$ cm for the CERN Super Proton Synchrotron (SPS) or Large Hadron Collider (LHC) cases. This length places an upper limit on the plasma densities and corresponding accelerating fields that can be driven by these single bunches, i.e., $E_{wb} =$ 37–75 MV/m when $\sigma_z/(c/\omega_{pe}) \simeq \sqrt{2}$. The self-modulation (SM) mechanism was proposed [8] to drive accelerating fields with amplitude 1 GV/m with a long proton bunch. Under the action of the transverse component of the wakefields, the bunch is periodically focused and defocused [9] and is transformed into a train of short bunches separated by the plasma wavelength $\lambda_{pe} = 2\pi c/\omega_{pe}$ with defocused protons in between. The bunch train resonantly drives wakefields to large amplitudes. In order to avoid transverse filamentation [10] and for the SM to fully develop [11], the bunch transverse size must be such that $\sigma_r/(c/\omega_{pe}) \leq 1$. With this choice of plasma density, $\sigma_z/(c/\omega_{pe}) \gg 1$ and the bunch drives wakefields with a period much shorter than $\sigma_z/c$ as demonstrated experimentally [12]. The plasma density is then $(\sigma_z/\sigma_r)^2$, and $E_{wb}$ is $\sigma_z/\sigma_r$ larger than with $\sigma_z/(c/\omega_{pe}) \cong \sqrt{2}$. Microbunching of a 22 MeV electron beam was observed recently [13].

The SM process can start from noise wakefields present in the plasma and from density modulation in the long bunch [14]. The SM can also be seeded to become a seeded-self-modulation (SSM) process [15]. Seeding is needed to control the result of the SM process and to deterministically inject electrons (or positrons) in the accelerating and focusing phase of the wakefields. Seeding methods include driving wakefields with a preceding laser pulse or particle bunch, shaping the drive bunch [16], or using a relativistic ionization front [15].

In this Letter, we demonstrate for the first time that self-modulation of a relativistic proton bunch leads to the formation of a train of many ($> 20$) microbunches (full modulation) separated by $\lambda_{pe}$, and that the process can be seeded with a relativistic ionization front. We show that, for over one order of magnitude in plasma density, the modulation frequency is equal to the plasma frequency as expected from theory [8]. We show that self-modulation occurs at least as far as one $\sigma_z$ behind the seed point, where the wakefields are expected to reach their maximum acceleration value [17].

In the Advanced Wakefield Experiment (AWAKE), the SPS proton bunch [400 GeV/proton, $(0.5-3)\times 10^{11}$ particles] is used to demonstrate the SSM process. The experimental setup is shown in Fig. 1 [18].

The proton beam is focused to a radius $\sigma_r \sim 200$ $\mu$m near the plasma entrance, corresponding to a nominal plasma density of $7\times 10^{14}$ cm$^{-3}$ ($\sigma_r/(c/\omega_{pe}) \simeq 1$). The plasma density can be adjusted between $0.5\times 10^{14}$ and $10.5\times 10^{14}$ cm$^{-3}$ [20]. The proton bunch and the laser pulse enter the 10-m-long vapor source colinearly. The laser pulse ($\sigma_{\mathrm{Laser}} = 120$ fs, $> 10(\mathrm{TW/cm^2})$) singly ionizes a Rubidium vapor over the 10 m source length [21], while copropagating within the bunch. It creates a sharp, relativistic ionization front with transition time $< \sigma_{\mathrm{Laser}} \ll (1/\omega_{pe})$, i.e., much shorter than the expected wakefield period (3–10 ps), thus effectively seeding wakefields. The density of the vapor, and indirectly of the plasma, is measured with white light interferometry with an accuracy of $\leq 0.5\%$ [22] and is constant over the length of the source, unless specified otherwise. The plasma radius [23] is ∼1 mm. Alignment between proton and laser beams (and thus plasma) is obtained and maintained using beam position monitors and a virtual laser line. The SSM process develops along the plasma with saturation expected before its end [24]. At saturation the microbunches are fully formed and propagate towards the plasma end and diagnostics. The modulated bunch exits the plasma and the laser pulse is blocked by a thin foil to protect downstream diagnostics. At the diagnostic, the bunch travels through a 280 μm Silicon wafer (OTR foil in Fig. 1) coated with 1 μm mirror-finished Aluminum. Protons emit (incoherent) backwards optical transition radiation (OTR) when crossing the vacuum-Aluminum boundary. The prompt OTR [25] has the same time and spatial structure as the bunch charge distribution at the wafer. The OTR is transported and imaged onto the entrance slit of a streak camera [26]. The slit selects a 74–110 μm wide transverse slice of the bunch, centered on the beam axis. The streak camera creates a time-resolved image of the OTR passing through the slit within 1.1 ns to 73 ps time windows and a 1 ps time resolution on the shortest timescale. The dimension along the slit gives the bunch transverse charge distribution at the wafer. Streak camera images shown in Figs. 2 and 3 are therefore time resolved images of the proton distribution in a thin slice. The wafer is placed 3.5 m after the plasma end and the streak camera observes the beam after this free space propagation [27]. Without plasma, the beam transverse size at the wafer is ∼560 μm.

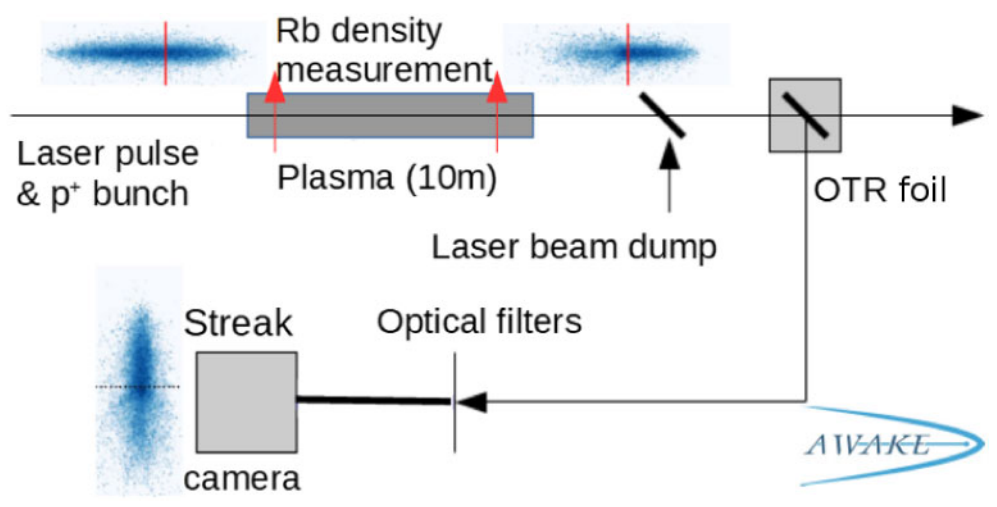

FIG. 1. The protons together with the laser pulse (red marker in the depicted proton bunches) enter the plasma. The first half of the bunch propagates in neutral Rubidium while the second half propagates in plasma. After the plasma, the laser hits a beam dump and the modulated bunch hits a foil where optical transition radiation (OTR) is created. The light is imaged onto the slit of a streak camera and is streaked to obtain a time resolved image. The optical filters limit the bandwidth of the light to ensure time resolution.

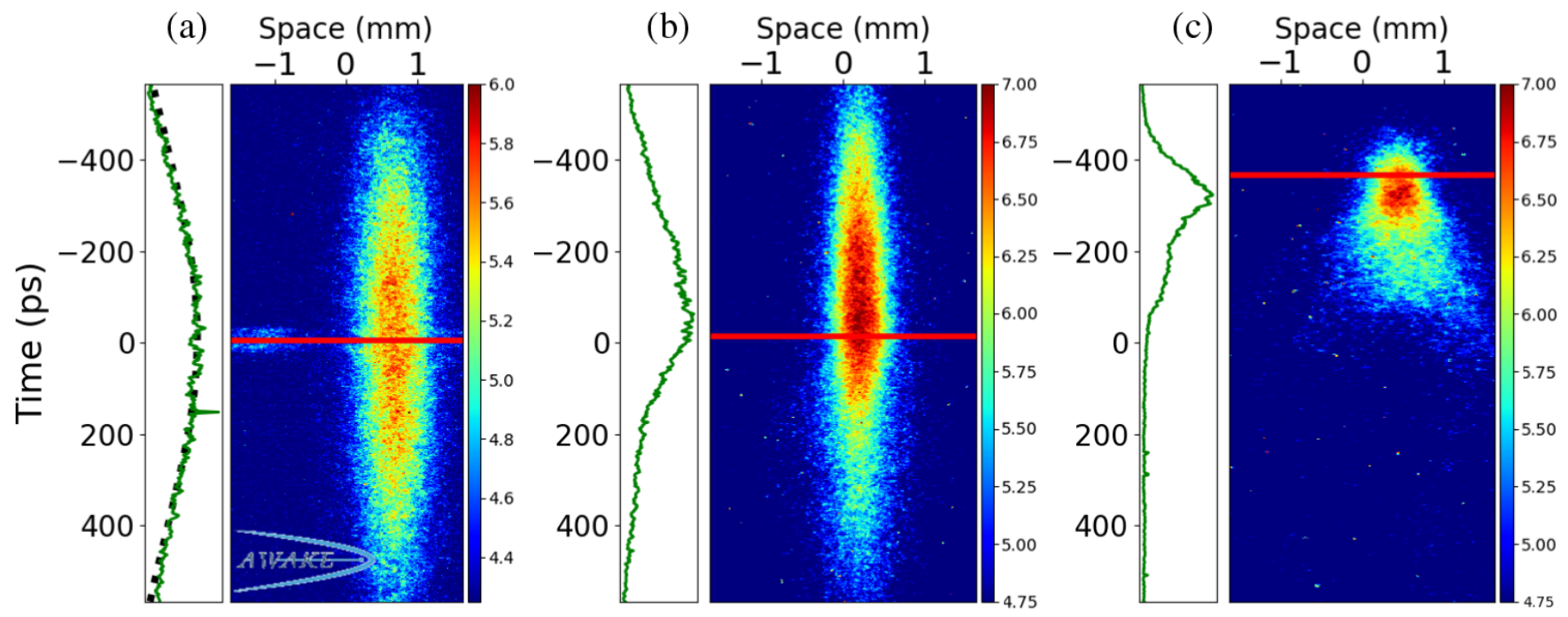

FIG. 2. (a) Streak camera image of the proton bunch without plasma. The bunch is on the right hand side of the image and the short laser pulse on the left hand side (see text). The green temporal profile is fitted with a Gaussian function (dotted black line, $\sigma_t = 437$ ps, $1.5 \times 10^{11}$ protons). The laser timing is marked by the red line ($t = 0$). (b) Image with plasma ($n_{\mathrm{Rb}} = 2.092 \times 10^{14}$ cm$^{-3}$, $3 \times 10^{11}$ protons) and ionizing laser pulse (blocked, not visible) placed as in (a). The effect of the plasma ($t > 0$) is visible in the image and on the bunch profile (green line). (c) Image as in (b), but with $n_{\mathrm{Rb}} = 2.558 \times 10^{14}$ cm$^{-3}$ and the ionizing laser pulse in the front of the bunch at −390 ps. The SSM effect and streak camera setting are such that the bunch charge behind the laser pulse is not visible. The spatial dimension displayed is that at the OTR wafer.

Figure 2(a) shows a streak camera image of the bunch in a 1.1 ns window, obtained without plasma. The green profile on the side of the image shows the time profile of the bunch obtained by summing along the image spatial dimension. The measured rms bunch duration is 437 ps as determined by Gaussian fit of the profile (dashed black line). No Rubidium vapor or high-power laser pulse were present. Images with only vapor or an ionizing laser pulse show no measurable differences. The timing of the ionizing laser pulse with respect to the bunch (red line) is obtained by allowing the laser pulse in low power mode to reach the streak camera following the same path as the OTR. Figure 2(a) shows this case with the laser pulse placed in the middle (in time) of the bunch. The laser pulse [much shorter than it appears in Fig. 2(a)] is displaced laterally to be visible in the image.

Figure 2(b) shows an image with the same timescale, but its temporal position with vapor and an ionizing laser pulse present. The high-power laser pulse is blocked and not visible, but is marked by the red line. Only the second half of the bunch [$t > 0$ ps in Fig. 2(b)] propagates in, and can interact with the plasma. The image shows that the back of the bunch is strongly affected by the plasma. Protons are defocused by the SSM, OTR is spread along and across the streak camera slit, and much less light is collected. However, significant charge remains along the bunch axis. We show below (Fig. 3) that this charge is made of microbunches generated by the SSM. The time profile of the bunch (green line) also shows the effect starting at the ionizing laser pulse time. However, on this (and the following) image, the time resolution of the streak camera is too low (∼12 ps) to resolve the expected SSM period (∼8 ps for $n_{pe} = 2.092 \times 10^{14}$ cm$^{-3}$). Therefore the microbunch structure is not visible.

In Fig. 2(c), the ionizing laser pulse is located 390 ps earlier than in Fig. 2(b), and, as expected, the effect also appears at that time. Unlike in Fig. 2(b) the streak camera settings do not allow the remaining charge along the bunch to be seen. Similar images were obtained with various Rubidium vapor densities in the $(0.5\text{–}10.5) \times 10^{14}\ \mathrm{cm}^{-3}$ range, with various laser pulse and bunch parameters. All showed a similar effect. These measurements show that it is possible to seed the SM with the relativistic ionization front.

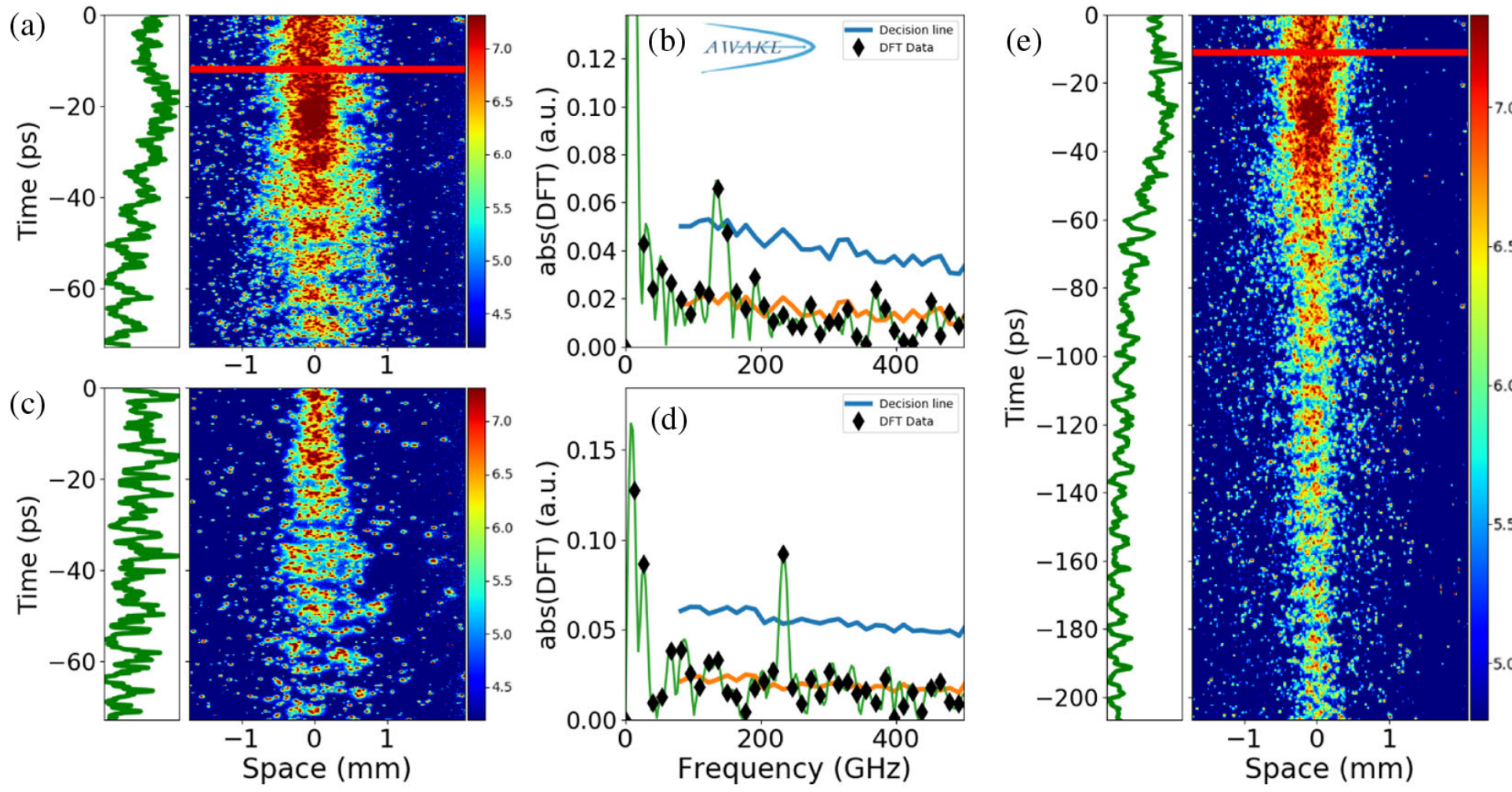

FIG. 3. Streak camera images on the fast (73 ps) timescale (a) at low ($n_{\mathrm{Rb}} = 2.457 \times 10^{14}\ \mathrm{cm}^{-3}$) and (c) at high ($n_{\mathrm{Rb}} = 6.994 \times 10^{14}\ \mathrm{cm}^{-3}$) plasma densities. Profiles obtained by summing the images along the spatial axis from −0.4 to 0.6 mm are displayed on the left-hand side of each image. The profile of image (a) shows the defocusing effect of SSM starting at the laser pulse time (∼10 ps). Image (c) is obtained ∼10 ps behind the ionizing laser pulse that is placed in the middle of the bunch as in Fig. 2. It is also obtained with a narrower band-pass filter (25 nm) than for image (a) and (e) (50 nm) to reduce the intensity of the light and decrease time resolution limitations originating from a broad OTR spectrum reaching the streak photocathode [26]. Figures (b) and (d) show the DFT power spectrum for the two profiles (black diamonds, no padding) as well as for background images (orange lines). The green lines depict the interpolated power spectrum (with padding). The blue lines show a noise threshold used for automatically detecting frequencies. Image (e) shows a low density case ($n_{\mathrm{Rb}} = 2.190 \times 10^{14}\ \mathrm{cm}^{-3}$) where the full train of microbunches is shown. The Rubidium (and thus plasma) density for image (e) has an upwards density gradient of 3.4%/10 m.

In order to observe the microbunches, we acquired images with the 73 ps (∼1 ps resolution) streak camera window, sufficient to visually resolve microbunches at low plasma densities ($\leq 5 \times 10^{14}\ \mathrm{cm}^{-3}$) and to detect the charge modulation at higher densities using discrete Fourier transform (DFT) [28]. Figure 3(a) shows such an image taken at $n_{\mathrm{Rb}} = 2.457 \times 10^{14}\ \mathrm{cm}^{-3}$. Starting at the top of the image, the presence of microbunches is visible over the whole image. The time profile (green line) also shows the microbunches and their periodicity. The periodicity is estimated at ∼7 ps. Figure 3(b) shows the DFT power spectrum of the image profile. With the window of 73 ps in 512 pixels, the DFT bin discretization is ∼14 GHz. The noise discrimination line (blue) is used to select only peaks that are unlikely to originate from noise (<1% probability of a noise peak above this line). The spectrum exhibit a clear peak above that level at 137 GHz. The frequency resolution of the DFT is not as precise as the density measurement ($\leq 0.5\%$). We interpolate the DFT bins by zero-padding the time domain profile to decrease the DFT discretization [29]. For the 73 ps rectangular window, half of the 3 dB bandwidth of the interpolation kernel function is 4 GHz, which we consider as the resolution limit for the frequency determination with interpolation. The zero padding factor used is ten, corresponding to a 1 GHz discretization. The interpolated spectrum has its peak at $(138 \pm 4)$ GHz, corresponding to a period of $(7.2 \pm 0.2)$ ps. This value is consistent with the expected plasma frequency of 141 GHz.

Figure 3(c) shows a similar streak camera image with $n_{\mathrm{Rb}} = 6.994 \times 10^{14}\ \mathrm{cm}^{-3}$. In this case, the modulation is not as clearly visible as at low density [Fig. 3(a)]. However, as we have previously shown, by using gated and beating laser beams imitating the OTR time structure expected from the modulated proton bunch, Fourier analysis can detect periods down to 2.2 ps [28], i.e., modulation frequencies up to 450 GHz even when the periodicity is not directly visible on the image. Figure 3(d) shows a peak in the DFT power spectrum at $(238 \pm 4)$ GHz, close to the 237 GHz plasma frequency expected at this density.

Short timescale images, as in Fig. 3, show the expected microbunches periodicity at times close to the seed point. Figure 3(e), obtained with a 206 ps window, a shortened proton bunch $\sigma_t \sim 240$ ps, and $n_{\mathrm{Rb}} = 2.190 \times 10^{14}\ \mathrm{cm}^{-3}$, shows that the modulation extends to the window limit, i.e., $\sim\sigma_z$, behind the seed point (red line). According to the Rubidium density measurement, the plasma frequency is 133 GHz and the interpolated microbunch frequency is $(133 \pm 2)$ GHz (3 dB bandwidth, 206 ps rectangular window). We note here that Fig. 3(a) shows defocused protons between microbunches, consistent with the fact that the microbunching is the result of a transverse and not of a longitudinal effect. This is confirmed by calculating the dephasing length between particles, including the highest possible energy gain or loss between protons ($E_{wb} = 3.1$ GV/m over 10 m, $n_{pe} = 10.5 \times 10^{14}$ cm$^{-3}$), yielding $\Delta L \cong 0.01\lambda_{pe}$. Figure 4 shows the result of modulation frequency measurements for seeded events with $1.3 \times 10^{14}$ cm$^{-3} \le n_{\mathrm{Rb}} \le 10.5 \times 10^{14}$ cm$^{-3}$. The measured modulation frequency is expected to be equal to the plasma frequency: $f_{\mathrm{mod}} = f_{pe} = (1/2\pi)\sqrt{(e^2/m_e\epsilon_0)}$ $n_{pe}^{1/2} \sim 89.8\beta^{1/2} n_{\mathrm{Rb}}^{1/2}$, where $\beta$ is the Rubidium fractional ionization. We perform a Bayesian fit (see Supplemental Material SM [30]) of the mean detected frequencies to determine $\alpha$ and $\beta$ values. The result shows that the mode values are $\alpha = 0.497$ and $\beta = 0.999$, consistent with the expected dependency $f_{\mathrm{mod}} = f_{pe}$ ($\alpha = 0.5$) and full ionization of the Rubidium vapor at all densities ($\beta = 1.0$). Similar results were obtained with seeding times earlier along the bunch.

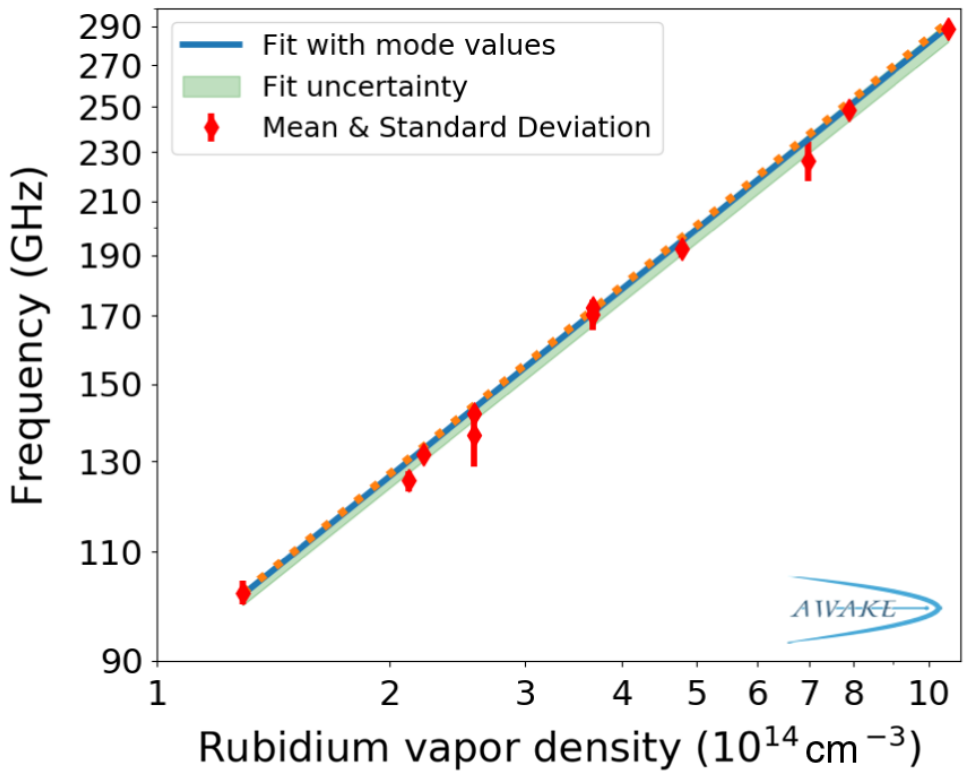

FIG. 4. Measured mean modulation frequencies (red dots, with standard deviation of the measured frequencies) as a function of the Rb vapor density $n_{\mathrm{Rb}}$ on a log-log plot. The orange dotted line corresponds to the expected $\alpha = 0.5$, $\beta = 1$ dependency of the modulation frequency. The mode values of the posterior distribution are plotted as the blue line.

Observing the seeded self-modulation is key for the acceleration of externally injected electrons in the wakefields driven by long proton bunches [31]. The results presented here with a relativistic ionization front (as opposed to a sharp rising charge distribution as in Refs. [12,13]) show for the first time the seeding and the development of the self-modulation of a long charged particle bunch in a plasma. At long timescales, we observe that the SSM effect starts at the time of the ionizing laser pulse and thus of the ionization front. At low plasma densities, fast timescale images show the formation of microbunches. They also show that protons in between the microbunches are defocused, which is consistent with the action of the transverse wakefields creating the microbunches. Systematic measurements of the modulation frequency versus Rubidium density show that the modulation frequency is equal to the plasma frequency. Future plans for using the seeded bunch are experiments for narrow energy spread ($< 1\%$) accelerated electrons and possible electron or positron collision applications [32].

The support of the Max Planck Society is gratefully acknowledged. This work was supported in parts by the Siberian Branch of the Russian Academy of Science (Project No. 0305-2017-0021), a Leverhulme Trust Research Project Grant RPG-2017-143 and by STFC (AWAKE-UK, Cockroft Institute core and UCL consolidated grants), United Kingdom; a Deutsche Forschungsgemeinschaft project grant PU 213-6/1 "Three-dimensional quasi-static simulations of beam self-modulation for plasma wakefield acceleration"; the National Research Foundation of Korea (No. NRF-2015R1D1A1A01061074 and No. NRF-2016R1A5A1013277); the Portuguese FCT—Foundation for Science and Technology, through Grants No. CERN/FIS-TEC/0032/2017, No. PTDC-FIS-PLA-2940-2014, No. UID/FIS/50010/2013 and No. SFRH/IF/01635/2015; NSERC and CNRC for TRIUMF's contribution; and the Research Council of Norway. M. Wing acknowledges the support of the Alexander von Humboldt Stiftung and DESY, Hamburg. The AWAKE collaboration acknowledge the SPS team for their excellent proton delivery.

*rieger@mpp.mpg.de